\documentclass[12pt,epsf]{article}

\usepackage{epsfig}

\def\be{\begin{equation}}
\def\ee{\end{equation}}
\def\beq{\begin{eqnarray}}
\def\eeq{\end{eqnarray}}
\def\e{\epsilon}

\def\N{{\cal N}}
\def\R{{\cal R}}
\def\RR{{\bf R^3}}
\def\({\left (}
\def\){\right )}
\def\[{\left [}
\def\[{\right ]}

\begin{document}

\begin{titlepage}
\bigskip
\rightline{}
\rightline{hep-th/0312123}
\bigskip\bigskip\bigskip\bigskip
\centerline{\Large \bf {Creating Naked Singularities and Negative 
Energy\footnote{To appear in the proceedings of the Nobel Symposium on
Cosmology and String Theory; Sigtuna, Sweden, August 14-19, 2003.}}}
\bigskip
     \bigskip\bigskip
           \bigskip\bigskip

 \centerline{\large  Gary T. Horowitz}
    \bigskip\bigskip
  \centerline{\em  Department of Physics, UCSB, Santa Barbara, CA 93106}
   \centerline{gary@physics.ucsb.edu }

\begin{abstract}
A brief review is given of three recent results concerning classical
solutions of gravitational theories: (1) With asymptotically
anti de Sitter boundary conditions, there are matter theories satisfying
the positive energy theorem which violate cosmic censorship. (2) Despite
supersymmetry, there
are solutions to $\N=8$ supergravity in which the total gravitational
energy is arbitrarily negative.
This theory can also violate cosmic censorship.
(3) A large class of supersymmetric compactifications (including all 
simply connected Calabi-Yau manifolds) have solutions with negative
four dimensional effective energy density.
\end{abstract}

\end{titlepage}

\baselineskip=18pt

\setcounter{equation}{0}
\section{Introduction}

I would like to review some surprising results concerning classical solutions
of gravitational theories. Since many of these theories arise as the
low energy limit of string theory, these results have direct implications for 
string theory and suggest a new approach for studying cosmological 
singularities in a quantum theory of gravity.
I will only have time to describe the main ideas and summarize the results.
This work was all done in collaboration with T. Hertog and K. Maeda, and
the reader  is referred to the original papers for more details 
\cite{Hertog:2003zs,Hertog:2003xg,Hertog:2003ru}.

Between
1965 and 1970, Hawking and Penrose proved a series of powerful theorems
showing that large classes of solutions to Einstein's equation are singular
\cite{Hawking:1969sw}.
These singularity theorems marked a major advance in our understanding
of strong gravitational fields.
However, they say nothing about event horizons. In particular, 
they do not prove that
black holes must form in gravitational collapse. It is possible that, in some
cases, one forms singularities that are visible to distant observers. These
are known as {\it naked singularities}.
In 1969, Penrose suggested that there might
be a ``cosmic censor" that forbid naked singularities from forming
\cite{Penrose:pc}. 
This has become known as cosmic censorship. I should emphasize that
cosmic censorship is not concerned with static timelike singularities like
the one in the negative mass Schwarzschild solution. Those singularities
are indeed naked, but they are present for all time. They are part of the
initial conditions. Cosmic censorship deals with singularities arising
from nonsingular initial data.

Cosmic censorship is clearly important for our understanding of black holes.
In fact, our entire theory of black holes
is based on the assumption that there are no naked singularities outside
the event horizon. 
But despite extensive work over the past three decades, we are still very far
from a complete proof \cite{Wald:1997wa}.
When something is hard to prove, it is often fruitful
to look for counter-examples, and many people have done so. It was shown 
early on that if you model matter by pressureless dust, then it is easy to
produce naked singularities \cite{Joshi:zg}.
One can take a spherical ball of dust and
start the outer shells collapsing inward faster than the inner shells. When
the shells cross, the density diverges and one gets a curvature singularity
which can lie outside the event horizon. But this is clearly an artifact
of the unphysical model of matter. Real matter has pressure. A clear signal
of the unphysical nature of this example
is that if one turns off gravity and just studies dust in flat space,
one has the same type of singularities. Since we are interested in singularities
arising from gravitational collapse, we should only consider matter which 
is nonsingular when evolved in Minkowski space, such as a scalar
field.

About ten years ago, Choptuik showed that spherically symmetric scalar
fields coupled to gravity can produce a naked singularity \cite{Choptuik:jv}.
This attracted
a lot of attention (and caused Stephen Hawking to lose a bet with Kip Thorne)
but it did not really disprove cosmic censorship. The reason is that 
Choptuik had to fine tune his initial data. Nearby initial data either
forms small black holes or results in the scalar field scattering and never
forming a singularity. Cosmic censorship deals with generic initial data.
To violate it, one needs to have an open set of initial data which 
evolve to form naked singularities.

The above examples assume the usual boundary conditions for an isolated
system: the spacetime is asymptotically flat at infinity. However there
has recently been a great deal of interest in spacetimes with nonzero
cosmological constant. When the cosmological constant is positive
(as suggested by the astrophysical data) and the matter Lagrangian
includes dilatons (as suggested by string theory) there is an
interesting class of counterexamples to cosmic censorship which
has received less attention \cite{Maeda:1998hs}. The action can
be taken to be
\be
S= \int [R - 2(\nabla\phi)^2 - e^{-2\phi}F^2 - 2\Lambda +{\cal L}_m]\sqrt{-g}
\ d^4 x
\ee
where $F$ is a Maxwell field,  $\Lambda >0$, and ${\cal L}_m$ describes
some charged matter.  Cosmic censorship can be violated since
there are
no static charged black hole solutions in this theory \cite{Poletti:1994ww}.
Roughly speaking the reason for this is that a static black hole in de Sitter
space is expected to have at least two horizons; a cosmological horizon
as well as a black hole horizon. On a static surface, the scalar field $\phi$
should reach both a local maximum  and a local minimum.
But its field equation $\nabla^2 \phi = -{1\over 2}e^{-2\phi}F^2$ 
only allows $\phi$
to have 
maximum or minimum values (but not both) whenever $F^2 \ne 0$. One can 
construct data in which the charged matter collapses and
forms singularities, but since there are no static black holes,
the singularities must be naked.

I will focus on the case of negative cosmological constant. I realize
that this is the wrong sign as far as cosmologists are concerned, but
in terms of understanding string theory, we are in much better shape for 
several reasons. Most importantly, we have a complete nonperturbative 
formulation of the theory provided by the AdS/CFT correspondence
\cite{Maldacena:1997re}. AdS refers to anti de Sitter space, 
the maximally symmetric
solution with negative cosmological constant, and CFT corresponds to 
a conformally invariant quantum field theory.
The remarkable claim (which is supported by a large
body of evidence \cite{Aharony:1999ti})
is that string theory with asymptotically AdS boundary
conditions is completely equivalent to a CFT.

It turns out that cosmic censorship is much
easier to violate in asymptotically AdS spacetimes than asymptotically 
flat spacetimes. This is a result of two facts:

\centerline{1) Black holes are harder to form.}

\centerline{2) Singularities are easier to form.}

\noindent
The first just follows from the form of the solution for a black hole
in AdS
\be\label{sads}
ds^2 = -\(1-{2M\over r}+{r^2\over \ell^2}\) dt^2 + \(1-{2M\over r}+
{r^2\over \ell^2}\)^{-1} dr^2 + r^2 d\Omega
\ee
where $\ell$ is the radius of curvature of AdS and $M$ is the mass of the
black hole. (If $M=0$, the metric is just AdS.)
This shows that the mass needed to form a black hole
of size $R_s \gg \ell$ is $M\propto R_s^3$. This is much larger than the
mass needed to form the same size black hole in asymptotically flat space
which is $M=R_s/2$.

The second fact is a result of a qualitatively new way to form singularities
with a scalar field in AdS. In asymptotically flat spacetimes, the main way to
form a singularity is to arrange the scalar field so that there is a large
energy density inside a small volume. This will produce a singularity, but
it also will usually produce a black hole.
Now consider a potential $V(\phi)$ for a scalar field with a negative minimum. 
Suppose we take  initial data in which the scalar field is simply
a constant $\phi_0$ (with $V(\phi_0)<0$)
and $\dot \phi =0$. By homogeneity, when this data is
evolved,
the metric can be written in Robertson-Walker form
\be
ds^2 = -dt^2 + a^2(t) d\sigma_{-1}
\ee
where $d\sigma_{-1}$ denotes the metric on a unit hyperboloid.
If $\phi_0$ is at the minimum of the potential, it
will stay there for all time and the solution for the scale factor
is $a(t) = \ell\cos(t/\ell)$ with $\ell^2 = -3/V(\phi_0)$.
This is just AdS in different coordinates than
(\ref{sads}). In particular,
$a(t)=0$ is just a coordinate singularity. But if $\phi_0$
is even slightly away from the minimum, the solution is dramatically changed.
Under evolution, the scalar field will start to oscillate. Einstein's equation
still forces the scale factor to vanish, but now when it vanishes, the
energy in the scalar field becomes infinitely blue-shifted and one has
a curvature singularity.

\setcounter{equation}{0}
\section{Cosmic censorship violation in AdS}

We will use the above two facts to show that there are theories of gravity
coupled to a scalar field $\phi$ which violate cosmic censorship 
\cite{Hertog:2003zs}.
Our strategy will be to find a potential $V(\phi)$ such that there is an
open set of initial data which forms a singularity in a large central 
region, but such that there is not enough mass to produce a black hole
big enough to enclose the singularity. For now we will not insist that
 $V$ comes from string theory, but ask if there is any potential which
will violate cosmic censorship. One physical condition that we will impose
is that $V$ satisfies a positive energy theorem. Recall that the total
energy is well defined for every asymptotically AdS spacetime \cite{Abbott82}.
The positive
energy theorem states that the total energy of all nonsingular
initial data is positive and vanishes if and only if the metric is
AdS everywhere \cite{Gibbons83,Boucher:1984yx,Townsend84}.
If this fails, the asymptotic AdS space is likely to be 
unstable, and the theory may not have a ground state.

We now show that
there indeed exist $V(\phi)$ such that the positive energy theorem holds
but cosmic censorship is violated.
The idea is to consider a potential like that shown in Fig. 1. $V(\phi)$
has a global minimum at $\phi=0$ and a local minimum at $\phi=\phi_1$,
both of which are negative. We will require that $\phi \to \phi_1$
asymptotically. One might worry that the positive energy theorem would
be violated since if one keeps $\phi$ very small inside a large ball
and then sends it over the barrier to $\phi_1$ in a thin shell,
it would appear that
the total energy is less than if $\phi=\phi_1$ everywhere. However this
intuition fails because the negative energy density causes the space to
be negatively curved, and a negatively curved hyperboloid has the property
that there is as much volume inside a shell at large radius  as there is
inside the entire ball of the same radius. When one computes the total 
energy, it turns out that
the positive energy theorem typically is satisfied if the barrier is large
enough, but is violated if the barrier is too small. Since we want to keep the
total mass small, we will adjust the height of the barrier to be close to
the transition point, but still satisfy the positive energy theorem.

\begin{figure}[htb]
\begin{picture}(0,0)
\put(0,80){V}
\put(120,0){$\phi$}
\put(190,80){$\phi_1$}
\end{picture}
\psfig{file=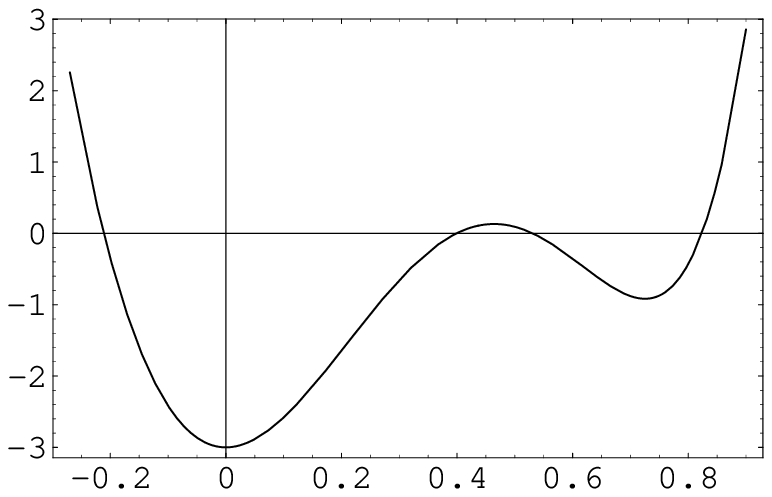,width=3.2in}
\caption{\sl A potential $V(\phi)$ that
satisfies the positive energy theorem for solutions that
asymptotically approach the local (AdS) minimum at $\phi_1$, but which
violates cosmic censorship.}
\label{2}
\end{figure} 

Now consider spherically symmetric initial data with $\phi(r)=
\e$ for $r<R_0$, $\phi=\phi_1$ for $r>R_1$ and continuously interpolating
in between. Here  $R_1 >R_0$ are two large radii. We also assume
that $\dot \phi =0$ so the initial data is time symmetric.
By spherical symmetry, we can assume the metric
takes the form
\be \label{metric}
ds^2 = \left(1-{2m(r)\over  r}+{r^2\over \ell^2}\right)^{-1} dr^2 +
r^2 d\Omega.
\ee
where now $\ell^{2} =-3/ V(\phi_1)$.
The mass function $m(r)$ is uniquely determined by the Einstein scalar
constraint
for any $\phi(r)$. The total mass is just the limit of $m(r)$ as
$r\to \infty$. Under evolution
we know that the region $r<R_0$ will become
singular in a time of order $V(0)^{-1/2}$. The question is whether this can
be hidden
inside a black hole.
If a black hole eventually forms 
we can trace the null geodesic generators back to our initial surface
where they will form a sphere of radius $R_s$ (see Fig. 2).
The black hole
area theorem still holds even with $V<0$, since it only requires the null
convergence condition $T_{\mu\nu} k^\mu k^\nu\ge 0$ for all null $k^\mu$.
So $R_{bh} \ge R_s$. We can now ask if there is enough mass to form
a black hole of size $R_s$. 

This is just a question about our initial data
and is easy to answer. It turns out that for some $\phi(r)$ the answer is
no! In fact, it is not even close. Recall that a large Schwarzschild-AdS
black hole has a mass $M\propto R_s^3$. This theory also has black holes
with scalar hair, but the positive energy theorem ensures that
a nontrivial scalar field outside the horizon only
increases the mass. In the extreme case where we adjust
the barrier to be right on the verge of violating the positive energy
theorem (but still satisfying it) and choose $\phi(r)$ to minimize the
mass, one finds that the mass available is $M \propto R_1$, but
the mass needed to form a black hole is $M_{bh} \propto R_1^3$ (since in
this case, $R_s \propto R_1$). Because of this large discrepancy, small
changes in the initial data will not effect this conclusion. One can
take nearby initial data which is not spherical or time symmetric and
still not have enough mass to form a black hole large enough to enclose
the singularity. The net result is that one has an open set of initial
data which produce naked singularities and cosmic censorship is violated in
this theory.

\begin{figure}[htb] 
     \centerline{\epsfxsize=9.6cm 
            {\epsfysize=5.5cm
    \epsffile{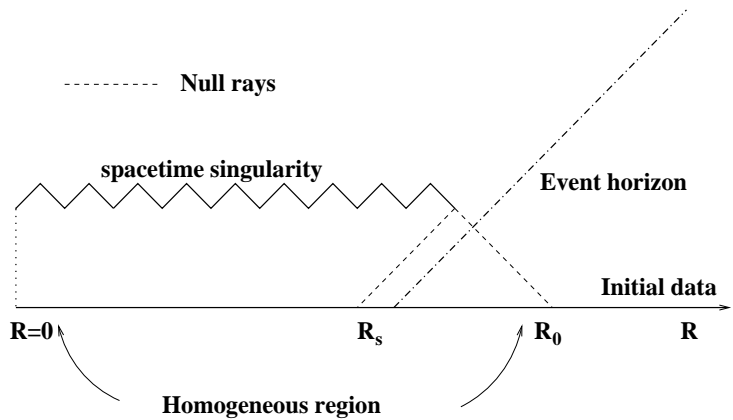}}}
   \caption{\sl If an event horizon encloses the singularity, it must have
     an initial size greater than $R_{s}$.}
   \label{4}
   \end{figure} 

Since we have gravitational collapse to a singularity which cannot
be enclosed inside a static black hole, we clearly have a violation
of the standard black hole paradigm. But is there really a naked singularity?
Inside the homogeneous region, the singularity is spacelike and cannot
be called naked. However,  we expect the singularity to extend outside
the region of homogeneous evolution. We don't know the exact solution here
but there are only a few possibilities. The most likely outcome is
that the singularity simply ends, in which case
the endpoint is naked. This includes the possibility that 
the singularity becomes timelike, because at
that point the classical evolution stops. (There is no unique evolution 
when the singularity is timelike since we don't know what boundary 
conditions to impose there.) Another possibility is that the
singularity remains spacelike and extends to infinity. This 
is a disaster much worse than a naked singularity, since then
localized initial data with  finite energy has created a big crunch 
singularity which  cuts off spacetime everywhere. We have not ruled
this out, but find it very unlikely. A final possibility is that the singularity
could become null. This is similar to a naked singularity in that
regions of arbitrarily large curvature are visible to distant observers.
Which of these possibilities actually occurs can be settled using numerical
relativity. This is currently being investigated \cite{garfinkle}.

\setcounter{equation}{0}
\section{Negative energy and naked singularities in supergravity}

We now try a similar construction with  a potential coming from
string theory. The low energy limit of string theory with
$AdS_5\times S^5$ boundary conditions is $D=5$, $\N=8$ supergravity. This 
theory has scalars with $m^2<0$. More than twenty years ago, Breitenlohner
and Freedman \cite{Breitenlohner:jf,Mezincescu85}
showed that these tachyons will not cause an instability in $AdS_d$
provided $m^2 \ge -(d-1)^2/4$ (where we have set the AdS radius to one).
This is known as the Breitenlohner-Freedman
(BF) bound. Supergravity has fields which saturate this bound. Since
our goal is to construct solutions with low total energy, it is natural to 
use these fields. The relevant potential $V(\phi)$ now has a  negative
maximum at $\phi=0$ and falls off exponentially for large $|\phi|$.
There are no local minima and no potential barrier. 

We want to consider initial data with $\phi=\phi_0$ for $r<R_0$ and
then $\phi\to 0$ asymptotically. The question is at what rate should
$\phi$ go to zero. We clearly want $\phi$ to go to zero as slowly as
possible. This will minimize the positive contribution to the energy
from the spatial gradients and maximize the negative contribution to the
energy coming from the fact that the potential is less than its
asymptotic value over a larger region. It turns out that the slowest
$\phi$ can fall off and keep the total mass finite is $1/r^2$. So we
take 
\be\label{indata}
\phi(r) = {A\over R^2_0}\ \  {\rm for} \ \ r<R_0, \qquad \phi(r) = {A\over 
r^2}\ \
{\rm for}\ \ r>R_0.
\ee
Assuming zero time derivatives, one can now solve the constraint
for the spatial metric and compute the
total mass. Surprisingly, it turns out  that it
is negative \cite{Hertog:2003xg}, $M\propto -A^2$!
Furthermore, by increasing $A$,
we can make the total mass arbitrarily negative. There is no lower bound.

How can this be? The theory is supersymmetric, and AdS is a supersymmetric
solution. There is supposed to be a positive energy theorem 
\cite{Gibbons83,Boucher:1984yx,Townsend84} which ensures that this cannot 
happen. Furthermore, the AdS/CFT
correspondence says that this theory is supposed to be equivalent to the
CFT which has a Hamiltonian bounded from below.

To resolve this contradiction,
let us review the positive energy theorems. The original Breitenlohner-Freedman
argument applied
to test fields satisfying the Klein-Gordon equation in AdS.
If $\xi^\mu$ denotes the global timelike Killing field,
the energy is just the integral of $T_{\mu\nu}\xi^\mu$ over a spacelike 
surface. This integral is not positive definite since $m^2 = -4$.
But if one sets $\phi = \psi/(1+r^2)$
one can integrate by parts and rewrite this energy as
\be\label{posenergy}
E= {1\over 2}\int\left [ (\dot \psi)^2 + (1+r^2) (D\psi)^2 + 4\psi^2
\right ] 
 {r^3\over (1+r^2)^3}  dr d\Omega_3 - \oint_\infty \psi^2 d\Omega_3
  \ee
The surface term vanishes if $\phi$ falls off faster than $1/r^2$, and
in this case the energy is manifestly positive. For all fields with $m^2$ above
the BF bound, finiteness of the energy requires the field to fall off faster
than $1/r^2$. However, for fields which saturate the bound, we have seen
that $1/r^2$ fall off is allowed by finite energy. The surface term is now
nonzero and negative causing $E<0$. Breitenlohner and Freedman
suggested that one should use
an ``improved" $T_{\mu\nu}$ which corresponds to adding a $\beta R\phi^2$
term to the Lagrangian. For a test field, this does not change the equation
of motion for $\phi$ since $R$ is a constant in AdS and acts like a mass term.
But in the full theory, this changes the gravitational dynamics. One
can check that $\N=8$ supergravity does not include these terms. So one
cannot use $\beta \ne 0$ to make the energy positive.


There is a complete nonlinear proof of the positive energy
theorem in AdS which is a generalization of the spinorial proof
of positive energy in asymptotically flat spacetimes originally found by
Witten \cite{Witten:mf}.
One solves a Dirac like equation for a spinor on a spacelike
surface $\gamma^i \hat \nabla_i \e =0$ where $\hat \nabla_i$ is a
supercovariant derivative (the derivative which appears in the supersymmetry
transformation laws). One can then derive an identity
in which  a surface term is equal to a manifestly positive volume integral.
If $\phi$ falls off faster than $1/r^2$, the surface term is the usual
total energy in AdS and one has a positive energy theorem. However, if
$\phi\sim 1/r^2$ asymptotically, there is another contribution to the
surface term at infinity coming from the asymptotic scalar field. (In a
test field limit, this extra contribution cancels the negative surface term
in (\ref{posenergy}).) The
sum of the two surface terms must be positive, but the usual energy need not
be. 

The net result is that the positive energy theorem requires boundary
conditions which are stronger than finite mass, and $M<0$ solutions exist.
However there is a modified energy (corresponding to the entire surface
term in the nonlinear proof described above) which is always positive.
This implies that AdS is stable: It cannot decay to another zero energy
solution. It also suggests that the CFT Hamiltonian should be identified
with this modified energy. This can probably be verified by computing
the supercharge, or more generally,
all the charges associated with the asymptotic superalgebra along the lines
of \cite{Henneaux85}. Why not call this modified energy the
``real energy" and forget the original definition? The answer is that 
the usual energy still governs the asymptotic behavior of the metric. Test
particles in the bulk solution at large radius will feel a negative
gravitational mass.

Having clarified the negative energy, we return to the question of evolution
of our initial data (\ref{indata}).
The central region collapses to a singularity.
If the total energy was conserved, then this singularity could not
 be hidden inside a black hole since the total energy is
negative and black holes must have positive mass. However it turns out
that for fields that saturate the BF bound and fall off like $A/r^2$
(and only in this case),
the energy is not automatically conserved. There can be a nonzero flux
of energy through infinity if the coefficient $A$ becomes time dependent.
If the energy grows sufficiently, a black hole
could form.

To control this, we can impose a large radius cut-off $R_1$ and require that
$\phi(R_1)=A/R_1^2$ is fixed. This is automatically
implemented in most numerical evolution schemes, and is a standard
regulator in discussions of AdS/CFT. However a  subtlety now
arises which can  be seen as follows. In the asymptotic AdS region, the
modes of a scalar field with $m^2=-4$
either fall off like $1/r^2$ or $\ln r/r^2$. In
the absence of a cut-off, only the
$1/r^2$ modes have finite energy and the $\ln r/r^2$ modes play no role.
However with a cut-off,  modes that
behave like $\ln r/r^2$ also have finite energy and cannot be ignored. It
turns out that these modes have energy which is even more negative than
the $1/r^2$ modes.
 The easiest way to
show that naked singularities can be produced is to modify our initial data.
We consider the following class of configurations,
$$\phi(r)=\phi_0 = {A \over \ln R_1}{\ln R_0  \over R_0^2} \qquad (r \le R_0)$$
\be
  \phi(r)={A \over \ln R_1}{\ln r \over r^2}
  \qquad ( R_0 < r < R_1)
  \ee
Since $\phi(R_1) \ne 0$, if a black hole forms, it must have some scalar 
hair. If one compares the mass of a static black hole with hair to our initial
data, one finds that the black hole has larger mass. Since the energy is
conserved in this regulated theory, the singularity must be naked.
Since we can perturb the initial data without changing the conclusion,
this yields generic
violation of cosmic censorship in supergravity.

This suggests a new approach to studying cosmological singularities
in string theory. We have
argued that our initial data will form a  singularity which is spacelike
for a while in AdS. But our initial data is time symmetric, so there is
a singularity in the past as well as the future. The solution thus
looks like a homogeneous cosmology with a big bang and big crunch singularity
embedded inside an asymptotically AdS spacetime (see Fig. 3).
What is the dual CFT
description? In this case, the CFT is $\N=4$ super Yang-Mills.
The operators dual to the fields which saturate the BF bound in
AdS correspond to the operators $Tr[X^i X^j - (1/6) \delta^{ij} X^2]$ 
where $X^i$ are the six scalars in the  super Yang-Mills theory.
(For a detailed
discussion of the relation between the bulk fields and the boundary operators
see \cite{Klebanov99,Bianchi:2001kw}.)  The large radius cut-off in AdS 
corresponds to a short distance cut-off in the gauge theory. It should
be possible to map our initial data into the gauge theory and study
its evolution.
There appears to be no reason for this evolution to stop.
This implies a string theory resolution of naked singularities. 
In principle, one should be able to use this to
determine if universes can bounce
through a cosmological singularity. The idea is to reconstruct the bulk
description of the evolved state in the gauge theory.
If the state corresponds to  a semiclassical metric which
is well defined everywhere shortly after the spacelike singularity,
it would show that universes can bounce in string theory (as often assumed
\cite{Gasperini:2002bn}). However, if it
is only well defined outside a
finite region, then there is no bounce. The
naked singularity continues in the bulk and the
cosmological singularity in the central region is truly an end of space and
time.

\begin{figure}[htb]
     \centerline{\epsfxsize=5.6cm  
            {\epsfysize=6.5cm
	                 \epsffile{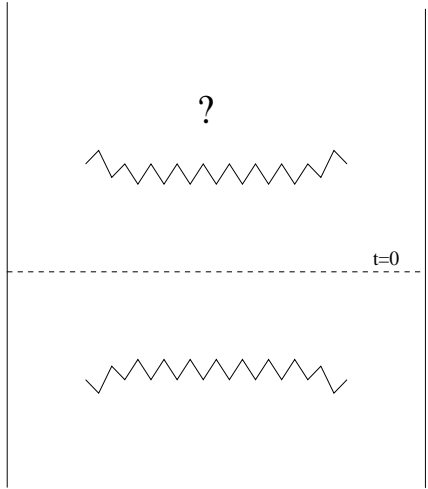}}}
   \caption{\sl Our solutions are like homogeneous universes beginning in a
    Big Bang singularity and ending in a Big Crunch, embedded in an 
   asymptotically anti de Sitter space. The dual field description can be used
   to study how the singularities are resolved.}
      \label{5}
      \end{figure}

\setcounter{equation}{0}
\section{Violation of cosmic censorship in asymptotically flat spacetimes?}

Suppose we add a constant to the potential in Fig. 1  so that the local minimum
at $\phi=\phi_1$ is at $V=0$. Then there are asymptotically flat solutions
and it is natural to ask if this theory also violates cosmic censorship.  
We still have the result that singularities
are easier to form since nearly homogeneous regions of $\phi$ rolling down
the negative part of the potential will still produce curvature singularities.
 But now black holes are not harder to
form since $M_{bh} \propto R_s$. So the outcome depends on more details of
the evolution. It should be straightforward to test whether
cosmic censorship is violated in this theory using  numerical relativity,
since one can start with spherically symmetric configurations.

It is natural to ask why we should consider
potentials with $V<0$. After all, they violate the dominate energy
condition. The answer is that they arise in many supersymmetric 
compactifications in string theory \cite{Hertog:2003ru}.

String theorists often consider spacetimes that approach $M_4\times K$
asymptotically where $M_4$ is four dimensional Minkowski spacetime and
$K$ is a compact Ricci flat space admitting a covariantly constant
spinor. Examples of $K$ include $T^n$, four dimensional K3, 
six dimensional Calabi-Yau
spaces, and seven dimensional manifolds with $G_2$ holonomy.
Let us ignore the other fields in string theory
and just consider vacuum solutions to general relativity in higher dimensions
compactified down to four dimensions. In other words, we are just
studying Kaluza-Klein theory.
Physically one is often interested in the four dimensional effective
description of these solutions. In four dimensions one has various matter
fields coming from the metric with various interactions. We claim that
some compactifications have configurations with negative four
dimensional energy density,
and in fact, this energy density can be arbitrarily negative. 

To justify this claim, lets begin with the following mathematical question:
Which compact manifolds $K$ admit Riemannian metrics of positive scalar
curvature? You might think that the answer is all of them since $\R >0$
is one scalar inequality on the entire metric. Indeed, if you change the sign,
this would be correct: All $K$ with dimension greater than two
admit metrics with
$\R<0$. But there is a topological obstruction to finding metrics with
$\R >0$ \cite{Lawson89}.
If you look at the standard examples above, $T^n$ and $K3$ do not admit any 
such metrics,
while all simply connected Calabi-Yau and $G_2$ manifolds do. There
is a theorem proved in 1990 \cite{Stolz90} which shows that in
dimensions greater than four, all simply 
connected manifolds
of dimension $3,5,6,7 \ {\rm mod}\ 8$ admit metrics with
$\R >0$. Note that this includes the cases of most interest to string
theory: six and seven.

What does this have to do with negative energy density?
Consider vacuum solutions
in higher dimension that approach $M_4\times K$ asymptotically. We can 
characterize this solution in terms of its initial data on a spacelike
surface $\RR\times K$. For time symmetric initial data, the constraint
equations reduce to  $\R=0$. If we take a product metric on  $\RR\times K$
with $\R_K >0$, then the only way to satisfy the constraint is to take
$\R_3<0$. But the usual $3+1$ constraint is $\R_3 = 16\pi\rho$,
so negative scalar
curvature is just like negative energy density. Since we can make $\R_K$
arbitrarily large by scaling $K$, we can make the energy density arbitrarily
negative.

Of course we cannot keep the metric a product everywhere since we need
the metric on $K$ to approach the standard Ricci flat metric at infinity.
But we can start with a large ball in $\RR$ and put on a product metric
with $\R_K >0$. Then we can have a transition region where the metric on 
$K$ approaches the Ricci flat one, and the metric on $\RR$ is adjusted to 
still satisfy the constraint. This shows that Calabi-Yau and $G_2$
compactifications have solutions with negative  energy density.
In other words, an effective four dimensional description would look 
qualitatively like Fig. 1 with a constant added so $V(\phi_1)=0$.
(This is only qualitative since a complete description would require
an infinite number of fields
in four dimensions.)

Despite having arbitrarily large regions  of arbitrarily
negative energy density, in this case the
total ADM energy always remains positive. This has been shown recently
 using a  Witten style 
positive energy theorem \cite{Dai03}.

\setcounter{equation}{0}
\section{Conclusion}

One can summarize the above discussion with two slogans: Supersymmetry
does not always imply positive energy, and positive energy does not
imply cosmic censorship. 

There are many open questions remaining. As we have just discussed, one
is whether cosmic censorship can be violated in asymptotically flat
spacetimes using potentials which are not positive definite.
Another is how common is cosmic censorship violation in supergravity. Are
there examples which do not involve negative energies?
Our examples of naked singularities
have all begun with (nearly) time symmetry initial data, so there is a 
singularity in the past as well as the future. While this may be
desirable for modeling the big bang, it would be of interest to show
that naked singularities could be produced in an evolution with no singularities
in the past. This seems likely, but has not been established.

I would like to conclude 
by pointing out that  if cosmic censorship is violated
in nature, it would not be a disaster. To the contrary, it would open up
the possibility  of directly observing the effects
of Planck scale or string scale curvature. This would be exciting development
for both observational astrophysicists and quantum gravity theorists.

\vskip .5 cm

\centerline{{\bf Acknowledgments}}
\vskip .5 cm
I want to thank my collaborators T. Hertog and K. Maeda for their assistance
on all aspects of the work described here. I also wish to thank the organizers
of the Nobel Symposium for a very stimulating conference.
This work was supported in part by NSF grant PHY-0244764.

\end{document}